\documentstyle[A4,12pt]{article}

\begin{document}

\begin{center}
{\Large On the Localization of One-Photon States}\\
\vspace{0.5truecm}
Corin~Adlard, E.R.~Pike and Sarben~Sarkar\\
\vspace{0.5truecm}
Department of Physics,\\
King's College London,\\
Strand, London WC2R 2LS,U.K.\\
\end{center}

\begin{abstract}

Single photon states with arbitrarily fast asymptotic power-law fall-off of 
energy density and photodetection rate are explicitly constructed. This
goes beyond the recently discovered tenth power-law of the Hellwarth-Nouchi
photon which itself superseded the long-standing seventh power-law of the
Amrein photon.

\end{abstract}

\newpage

\section{Introduction}

Given any solution of the source free Maxwell's equations it is possible to 
write down a corresponding one photon state
\begin{equation}
\label{1photonstate}
|\phi\rangle=\sum_{\lambda=\pm1}
\int\frac{d^3k}{(2\pi)^3}f_{(\lambda)}(\vec{k})
a_{(\lambda)}^{\dagger}(\vec{k})|0\rangle,
\end{equation}
where $|0\rangle$ represents the vacuum, and $\lambda$ denotes helicity. This 
is a general wavepacket state. The transverse nature of the classical solution
is reflected in the photon having only 2 (rather than 3) spin degrees of freedom
(i.e. 2 helicities). In particular we will show that
\begin{equation}
\label{fk-ak}
\sum_{\lambda}\vec{\varepsilon}_{(\lambda)}(\vec{k})f_{(\lambda)}(\vec{k})=
\sqrt{\frac{\varepsilon_0\omega_{\vec{k}}}{\hbar}}
\tilde{\vec{A}}(\vec{k}),
\end{equation}
where $\tilde{\vec{A}}(\vec{k})$ is the Fourier transform of $\vec{A}(t,
\vec{r})$, the classical solution for the vector potential, and $\vec{\varepsilon}_{(\lambda)}(
\vec{k})$ is the polarisation vector associated with helicity $\lambda$ and
wavevector $\vec{k}$. It has been known for a long time that the configuration
space representation of $|\phi\rangle$ cannot have delta function support or
support in a finite region of space \cite{Pike1995},\cite{Newton1949}. 
This prompted Jauch and Piron \cite{Jauch1967}
, and Amrein \cite{Amrein1969} to make a formal modification of the imprimitivity formulation of 
localisability of elementary particles to a generalised imprimitivity applicable to
photons. It was believed (without proof) that such a construction gave the tightest 
 isotropically localised single photon state. Such states can be shown to 
have an asymptotic ${1 \over {\left| {\vec r} \right|^7}}$ fall-off of its energy
density and photodetection rate \cite{Pike1987}. The situation remained thus for about a decade
until recently when a particular explicit solution of Maxwell's equation by 
Hellwarth and Nouchi led to a one-photon state with an asymptotic fall-off of
 ${1 \over {\left| {\vec r} \right|^{10}}}$  for the photodetection rate \cite{Sarkar1994} 
at any finite time $t$. Consequently it became
clear that the state from the generalised imprimitivity construction was not
the most isotropically localised possible. However the question of the
existence of a fundamental limit to the sharpness of the power-law fall-off
remained open. In this paper we will show that single photon states with
arbitrarily high powers of asymptotic fall-off can be explicitly constructed.

\section{The EDEPT solutions.}

Following the same calculations as in the paper of Ziolkowski \cite{Ziolkowski1989}, we can obtain 
the (focused) EDEPT (electromagnetic directed-energy pulse-train) solutions
 for the classical vector potential, expressed in cylindrical coordinates,
 as the real and imaginary parts of
\begin{equation}
\vec{A}(\tau,\rho,z)=
\frac{2\alpha\mu_0g_0^{\alpha}\vec{e}_{\theta}\rho
\left[g_1+i(z-\tau)\right]^{\alpha-1}}
{[r^2-\tau^2+i(g_2-g_1)z-i(g_2+g_1)\tau+g_1g_2]^{\alpha+1}},
\end{equation}
where $\alpha$ is an integer, $r^2=\rho^2+z^2$, $\tau =c t$  and, in the notation of Ziolkowski, 
we have set $b=0$, 
$\beta=g_0$, $z_0=g_1$ and $a\beta=g_2$. As we see from $(1)$ and $(2)$ these solutions provide a
convenient way of producing normalisable single photon states with all
the required transversality properties. 

Note that the leading power in distance in the denominator of the above is 
$r^{2\alpha+2}$ and that in the numerator the leading power is only  $\alpha$ at any finite time $t$.
Thus there exist real or imaginary  solutions for the vector  potential which
fall off asymptotically as $\frac{1}{r^{\alpha+2}}$ where $\alpha$ is odd for a
real solution and even for an imaginary solution. The solution of Hellwarth and
Nouchi is obtained from Eqn $(1)$ with $\alpha=1$. Since the electric and 
magnetic fields are the  derivatives of the vector potential with
respect to distance and time these fields will also inevitably fall off
asymptotically as $\frac{1}{r^{\alpha+c}}$. Here $c$ is some small integer whose 
value depends on whether a real or imaginary solution is required. Indeed an explicit
calculation shows that the electric and magnetic fields fall off asymptotically as a
power which increases linearly with increasing $\alpha$. These expressions are rather
long and so we do not give them here. Since we have not imposed an upper bound on
$\alpha$, other than it remain finite, we conclude that there is, in principle,
 no limit to the asymptotic rate of fall-off of the electric or magnetic energy 
densities of these {\em classical} solutions of Maxwell's equations.  

\section{Arbitrarily Localized One-photon states.}

We shall now demonstrate that there exist one-photon states with the same
electric and magnetic energy densities as the classical EDEPT solutions. 
Firstly, the Hamiltonian density of a free electromagnetic field in vacuum can
be written as
\begin{equation}
\label{quant_hdensity}
{\cal H}(t,\vec{r})=\frac{\varepsilon_0}{2}{\cal N}
\left(\vec{E}^2(t,\vec{r})+c^2\vec{B}^2(t,\vec{r})\right),
\end{equation}
where $\frac{1}{\varepsilon_0\mu_0}=c^2$ and where ${\cal N}$ represents 
normal operator ordering and $\vec{E}$ and $\vec{B}$ are 
free operator fields given, in the Coulomb gauge, by the usual relations 

\begin{equation}
\label{electric_def}
\vec{E}=-\frac{\partial\vec{A}}{\partial t}
\end{equation}
and
\begin{equation}
\label{magnetic_def}
\vec{B}=\vec{\nabla}\times\vec{A}.
\end{equation}
As usual, normal ordering is
required to remove the divergent energy density of the vacuum. 
The vector potential (operator) may be written, in vacuum as 
\begin{equation}
\label{vector_pot_def}
\vec{A}(t,\vec{r})=\sum_{\lambda=\pm1}
\int\frac{d^3k}{(2\pi)^3}
e^{-i(\omega_{\vec{k}}t-\vec{k}\cdot\vec{r})}
\sqrt{\frac{\hbar}{2\varepsilon_0\omega_{\vec{k}}}}
\vec{\varepsilon}_{(\lambda)}(\vec{k})a_{(\lambda)}(\vec{k})+\mbox{h.c.}
\end{equation}
For the general one photon state $|\phi\rangle$ the energy density is given by
\begin{eqnarray}
& &
\label{udensity}
u_{\mbox{qm}}(t,\vec{r})=\langle\phi|{\cal H}(t,\vec{r})|\phi\rangle
=\frac{\hbar}{2}\sum_{\lambda\lambda'}\int\frac{d^3kd^3k'}{(2\pi)^6}
f^*_{(\lambda)}(\vec{k})f_{(\lambda')}(\vec{k}')
\nonumber\\& &\times
e^{i(\omega_{\vec{k}}-\omega_{\vec{k}'})t-i(\vec{k}-\vec{k}')\cdot\vec{r}}
\sqrt{\omega_{\vec{k}}\omega_{\vec{k}'}}
\nonumber\\& &\times
\left(\vec{\varepsilon}^{\hspace{.1cm}*}_{(\lambda)}(\vec{k})
\cdot\vec{\varepsilon}_{(\lambda')}(\vec{k}')
+\hat{\vec{k}}\times\vec{\varepsilon}^{\hspace{.1cm}*}_{(\lambda)}(\vec{k})\cdot
\vec{\hat{k}}'\times\vec{\varepsilon}_{(\lambda')}(\vec{k}')\right).
\end{eqnarray}
Furthermore, we require a single-photon state to be normalizable and so

\begin{equation}
\label{Normalizability}
\langle\phi|\phi\rangle
=\sum_{\lambda}\int\frac{d^3k}{(2\pi)^3}\left
|f_{(\lambda)}(\vec{k})\right|^2<\infty.
\end{equation}

Now, since Ziolkowski uses a classical theory, we would like to find a 
one-photon state, $f_{(\lambda)}(\vec{k})$, such that it has the 
same energy density as the Hellwarth-Nouchi solutions. So we must find a suitable 
momentum-space expression for a classical energy density. The classical 
version of (\ref{quant_hdensity}) is the energy density
\begin{equation}
\label{class_hdensity}
u_{\mbox{cl}}(t,\vec{r})=\frac{\varepsilon_0}{2}
\left(\vec{E}^2(t,\vec{r})+c^2\vec{B}^2(t,\vec{r})\right),
\end{equation}
where $\vec{E}$ and $\vec{B}$ are now classical fields, so do not require 
normal ordering. Now, defining the Fourier transforms of the (real) electric 
field by
\begin{equation}
\label{eft}
\vec{E}(t,\vec{r})=
\int\frac{d^3k}{(2\pi)^3}\tilde{\vec{E}}(\vec{k})
e^{-i(\omega_{\vec{k}}t-\vec{k}\cdot\vec{r})}
=\int\frac{d^3k}{(2\pi)^3}\tilde{\vec{E}}^{*}(\vec{k})
e^{i(\omega_{\vec{k}}t-\vec{k}\cdot\vec{r})}
\end{equation}
and similarly for the magnetic field, and using these expressions in 
(\ref{class_hdensity}), we find
\begin{equation}
\label{ucl2}
u_{\mbox{cl}}(t,\vec{r})=\frac{\varepsilon_0}{2}\int\frac{d^3kd^3k'}{(2\pi)^6}
e^{i(\omega_{\vec{k}}-\omega_{\vec{k}'})t-i(\vec{k}-\vec{k}')\cdot\vec{r}}
\left(
\tilde{\vec{E}}^{\hspace{.1cm}*}(\vec{k})\cdot\tilde{\vec{E}}(\vec{k}')+
c^2\tilde{\vec{B}}^{\hspace{.1cm}*}(\vec{k})\cdot\tilde{\vec{B}}(\vec{k}')
\right).
\end{equation}
On substituting the definition of the Fourier transform of 
$\vec{A}(t,\vec{r})$
\begin{equation}
\label{aft}
\vec{A}(t,\vec{r})=\int\frac{d^3k}{(2\pi)^3}\tilde{\vec{A}}(\vec{k})
e^{-i(\omega_{\vec{k}}t-\vec{k}\cdot\vec{r})}
\end{equation}
into (\ref{electric_def}) and (\ref{magnetic_def}) we obtain
\begin{equation}
\label{ak-ek}
\tilde{\vec{E}}(\vec{k})=i\omega_{\vec{k}}\tilde{\vec{A}}(\vec{k})
\end{equation}
and
\begin{equation}
\label{ak-bk}
c\tilde{\vec{B}}(\vec{k})=
i\omega_{\vec{k}}\hat{\vec{k}}\times\tilde{\vec{A}}(\vec{k}),
\end{equation}
where we have written $\vec{k}=|\vec{k}|\hat{\vec{k}}$ and
$\omega_{\vec{k}}=c|\vec{k}|$. So, substituting (\ref{ak-ek}) and (\ref{ak-bk}) 
into (\ref{ucl2}) we find
\begin{eqnarray}
\label{ucl3}
u_{\mbox{cl}}(t,\vec{r})&=&
\frac{\varepsilon_0}{2}\int\frac{d^3kd^3k'}{(2\pi)^6}
\omega_{\vec{k}}\omega_{\vec{k}'}
e^{i(\omega_{\vec{k}}-\omega_{\vec{k}'})t-i(\vec{k}-\vec{k}')\cdot\vec{r}}
\nonumber\\& &\times
\left(\tilde{\vec{A}}^{\hspace{.1cm}*}(\vec{k})\cdot
\tilde{\vec{A}}(\vec{k}')+
\hat{\vec{k}}\times\tilde{\vec{A}}^{\hspace{.1cm}*}(\vec{k})
\cdot\hat{\vec{k}}'\times\tilde{\vec{A}}(\vec{k}')\right).
\end{eqnarray}
On using the Coulomb gauge condition in (\ref{aft}) we find in 
momentum space
\begin{equation}
\label{cgaugek}
\vec{k}\cdot\tilde{\vec{A}}(\vec{k})=0,
\end{equation}
that is, $\vec{A}(\vec{k})$ is always orthogonal to $\vec{k}$. Thus, we may 
expand $\tilde{\vec{A}}(\vec{k})$ on the basis of two orthonormal polarization 
vectors $\vec{\varepsilon}_{(\pm1)}(\vec{k})$ such that
\begin{equation}
\label{orthonormal}
\vec{\varepsilon}_{(\lambda)}(\vec{k})\cdot
\vec{\varepsilon}_{(\lambda')}(\vec{k})
=\delta_{\lambda\lambda'}
\end{equation}
and, of course
\begin{equation}
\vec{\varepsilon}_{(\lambda)}(\vec{k})\cdot\hat{\vec{k}}=0.
\end{equation}
So, we may write
\begin{equation}
\label{aexp}
\tilde{\vec{A}}(\vec{k})=
\sum_{\lambda=\pm1}\tilde{A}_{(\lambda)}(\vec{k})
\vec{\varepsilon}_{(\lambda)}(\vec{k}).
\end{equation}
Substituting this into (\ref{ucl3}) we find
\begin{eqnarray}
& &
\label{ucl4}
u_{\mbox{cl}}(t,\vec{r})=\frac{\varepsilon_0}{2}\sum_{\lambda\lambda'}
\int\frac{d^3kd^3k'}{(2\pi)^6}
\tilde{A}^*_{(\lambda)}(\vec{k})\tilde{A}_{(\lambda')}(\vec{k}')
\omega_{\vec{k}}\omega_{\vec{k}'}
e^{i(\omega_{\vec{k}}-\omega_{\vec{k}'})t-i(\vec{k}-\vec{k}')\cdot\vec{r}}
\nonumber\\& &\times
\left(\vec{\varepsilon}^{\hspace{.1truecm}*}_{(\lambda)}(\vec{k})
\cdot\vec{\varepsilon}_{(\lambda')}(\vec{k}')+
\hat{\vec{k}}\times\vec{\varepsilon}^{*\hspace{.1truecm}}_{(\lambda)}(\vec{k})
\cdot\hat{\vec{k}}'\times\vec{\varepsilon}_{(\lambda')}(\vec{k}')\right)
\end{eqnarray}

Thus, comparing (\ref{udensity}) and (\ref{ucl4}) we see that the classical and
quantum electric and magnetic energy densities will be equal if the 
single-photon state is given in momentum space by
\begin{equation}
\label{ak-fk}
f_{(\lambda)}(\vec{k})=\sqrt{\frac{\varepsilon_0\omega_{\vec{k}}}{\hbar}}
\tilde{A}_{(\lambda)}(\vec{k}).
\end{equation}

It can be shown for EDEPT solutions that such $f_{(\lambda)}(\vec{k})$ are normalisable.
 Thus, we have shown that there indeed exist single-photon states which have
an arbitrarily fast power law fall-off of their energy densities.

We shall now demonstrate that the detection rate for the localized 
one-photon states also fall off with an arbitrarily fast power-law. 
The detection rate of a single-photon state $|\phi\rangle$ for an ideal, 
point-like photon detector at $\vec{r}$ may be shown to be proportional to
\begin{equation}
S^{ij}\langle\phi|E^{(-)}_i(t,\vec{r})E^{(+)}_j(t,\vec{r})|\phi\rangle
\end{equation}
where $S^{ij}$ is the detector sensitivity, see for example Glauber
\cite{Glauber1964}. Let us note from the paper of Ziolkowski that the 
electric field of the doughnut solutions is polarized such that its classical 
electric field is in the $\vec{e}_{\theta}$ direction. It may similarly be 
shown that the `photodetection amplitude' 
$\langle0|\vec{E}^{(+)}(t,\vec{r})|\phi\rangle$ 
for the corresponding photon state $|\phi\rangle$ will be in this direction. 
Using this observation, we find that the detection rate is proportional to the 
electric energy density
\begin{equation}
\varepsilon_0
\langle\phi|\vec{E}^{(-)}(t,\vec{r})\cdot\vec{E}^{(+)}(t,\vec{r})|\phi\rangle.
\end{equation}
Now, since the electric energy densities of our 
single-photon states are equal to those of the corresponding EDEPT solutions, 
the single-photon detection rates are simply proportional to the 
classical electric energy densities, which fall off asymptotically with an
arbitrarily fast power-law. Hence, we have demonstrated that by choosing $\alpha$
the detection rates of these single-photon states can be made to have a power-law
 fall off which is arbitrarily rapid.

\vspace{0.5truecm}
{\bf Acknowledgement}
\vspace{0.5truecm}

One of us (C.A.) was supported by a Junior Research Studentship from the
School of  Physical Sciences and Engineering of King's College London.

\end{document}